\documentclass[useAMS,usenatbib,referee]{biom}

\newcommand{\bsl}{\mbox{\boldmath $\lambda$}}
\newcommand{\bsL}{\mbox{\boldmath $\Lambda$}}
\newcommand{\bsP}{\mbox{\boldmath $\Psi$}}

\newcommand{\bsS}{\mbox{\boldmath $\Sigma$}}
\newcommand{\bsth}{\mbox{\boldmath $\theta$}}

\newcommand{\bsPH}{\mbox{\boldmath $\Phi$}}

\newcommand{\bsOm}{\mbox{\boldmath $\Omega$}}

\newcommand{\mbf}{\mathbf{f}}

\newcommand{\mbx}{\mathbf{x}}

\newcommand{\mbl}{\mathbf{l}}
\newcommand{\mbC}{\mathbf{C}}
\newcommand{\diag}{\rm{diag}}
\newcommand{\mbe}{\mathbf{e}}

\usepackage[figuresright]{rotating}
\usepackage{color, amsmath}
\usepackage[normalem]{ulem}
\usepackage{caption}
\usepackage{hyperref}

\title[Multi-study Factor Analysis]{Multi-study Factor Analysis}

\author
{Roberta De Vito$^{1,*}$\email{rvito@princeton.edu},  Ruggero Bellio$^{2}$, 
 Lorenzo Trippa$^{3,4}$,
and Giovanni Parmigiani$^{3,4}$
\\$^1$Department of Computer Science, Princeton University, Princeton, NJ, USA
\\$^2$Department of Economics and Statistics, University of Udine, Udine, Italy
\\$^3$Department of Biostatistics and Computational Biology, Dana-Farber Cancer Institute, Boston, MA, USA
\\$^4$Department of Biostatistics, Harvard T. H. Chan School of Public Health, Boston, MA, USA}

\begin{document}







\begin{abstract}
We introduce a novel class of factor analysis methodologies for the joint analysis of multiple studies. The goal is to separately identify and estimate 1) common factors shared across multiple studies, and 2) study-specific factors.  We develop a fast Expectation Conditional-Maximization algorithm for parameter estimates and we provide a procedure for choosing the common and specific factor. We present simulations evaluating the performance of the method and we illustrate it by applying it to gene expression data in ovarian cancer. In both cases, we clarify the benefits of a joint analysis compared to the standard factor analysis.
We have provided a valuable tool to accelerate the pace at which we can combine unsupervised analysis across multiple studies, and understand the cross-study
reproducibility of signal in multivariate data. An R package, namely \textit{MSFA}, is implemented and is available on GitHub at \url{https://github.com/rdevito/MSFA}.
\end{abstract}


\maketitle

\section{Introduction \label{section intro}}

Analyses that integrate multiple sources, studies, and data-collection technologies are common in
 current statistical research. When considering multiple studies, a fundamental challenge is learning common features shared among studies while isolating the variation specific to each study.
Two important statistical questions remain largely unanswered in this context: i) To what extent is the common signal shared across studies? ii) How can  this shared signal be extracted?
In this paper we develop a methodology to address these two questions 
for   multi-study factor analysis. 

Joint factor analysis of multiple studies is often used in several areas of science.
For example \cite{scaramella2002} researched adolescent delinquent behavior in two independent samples analyzing the same variables in order to identify the shared patterns across those two different samples. \cite{andreasen2005} studied remission in schizophrenia, applying factor analysis (FA)  for each individual samples. The work showed that replicable results are found across all these FA leading to similar components.
In nutritional epidemiology \cite{edefonti2012} analyzed the diet habits and the risk of head and neck cancer. In such work five different populations shared exactly the same variables (e.g. nutrients) to be then merged in one single ``effective'' population. Then they 
 applied  FA  to  this effective population to determine common dietary patterns and their relation with head and neck cancer.   \citet{wang2011} used FA to obtain a unified gene expression measurement from distinct types of measurements on the same sample.

These examples illustrate the urgent need for a model able to handle multiple studies and to derive in a single analysis (1) factors that capture common information, shared across studies, and (2) study-specific factors.

Our own motivating example derives from the analysis of gene expression data \citep{irizarry2003,shi2006,kerr2007}. In gene expression analysis, as well as in much of high-throughput biology analyses on human populations, variation can arise from the intrinsic biological heterogeneity of the populations being studied, or from technological differences 
 in data acquisition. In turn both these types of variation can be shared across studies or not. As noted by \citet{Garrett-Mayer2008}, the fact that 
the determinants of both natural and technological variation differ
 across studies implies that study-specific effects occur in most
datasets. Both common and study-specific effects can be strong, and both need to be identified and studied.
Our interest in this issue is a natural development of our previous work on unsupervised identification of 
integrative correlation \citep{Parmigiani2004a,Garrett-Mayer2008,Cope:2014jo}, and multi-study supervised analyses including cross-study differential expression \citep{Scharpf:2009wy}, 
multi-study gene set analysis \citep{tyekucheva2011}, comparative meta-analysis \citep{Riester:2014ga,Waldron:to}, and cross study validation \citep{bernau2014}.

In high-throughput biology, as well as in a number of other areas of application, the ability to separately estimate common and study-specific factors can contribute significantly to two important questions: the cross-study validation question of whether factors are found repeatedly across multiple studies; and the meta-analytic question of more efficiently estimating the factors that are indeed common. With regard to interpretation, the shared signal is more likely to capture genuine biological information, while the study-specific signal can point to either artifactual or biological sources of variation. 
Thus, modeling both shared and unshared factors may enable a more reliable identification of artifacts, facilitate more efficient experimental designs, and inform further technological advances.

In this article we propose a dimension-reduction approach that allows for 
joint analysis of multiple studies, achieving the goal of capturing common factors. Specifically, we define 
a generalized version of FA, able to handle multiple studies simultaneously. Our model,
termed Multi-study Factor Analysis (MSFA), learns the common features shared among studies, and identifies the unique variation present in each study.

While unsupervised multi-study analysis is not an adequately studied field, our work draws from existing foundations from related problems. 
In the social science literature, there is extensive methodology
to identify factor structures shared among different groups, forming the body of  \emph{multigroup factor analysis} methods (see, among many others, \cite{thurstone1931, joreskog1971, meredith1993}). 
Such methods are mainly focused on investigating measurement invariance among different groups, that typically results in testing whether 
the data support the hypothesis of a common loading matrix
across groups. A notable special case is given by partial  measurement invariance \citep[see for example][]{byrne1989testing}, which inspired our mathematical formulation.
In our MSFA we have extended the scope to   detection of  both  study-specific  factors   and  factors  that  are  identical  across multiple   studies. Our MSFA has also an exploratory nature, different from the confirmative approach under which measurement invariance is usually investigated in the social sciences. 
 
The plan of the paper is as follows. Section 2 introduces the MSFA, and describes the estimation of model parameters based on maximum likelihood, implemented via an ECM algorithm. Section 3 presents simulation studies, providing numerical evidence on the performances of the proposed estimation methods. Next  the determination of 
 the dimension of the latent 
factor  is investigated,  tackled by casting it within the framework of model selection. Section 4 illustrates the application of the methodology
to study of the Immune System pathway in ovarian cancer. Section 5 contains the final discussion.

\section{Methods}

\subsection{The multi-study factor analysis (MSFA) model}
The methodology proposed  here  has two main goals.
First, we  combine multiple studies to identify  common factors   that  are  consistently observed across the studies. Second,  we  explicitly model and identify  additional components of variability  specific to single  studies,  through  study-specific factors.   
  The latter aims to identify variation that lacks cross-study reproducibility, and separate it from variation that does.

We consider $S$ studies, each with the same $P$ variables. Generic study $s$ has $n_s$ subjects and, for each subject,
 $P$-dimensional centered data vector $\mbx_{is}$ with $i=1,\dots,n_s$.
We begin by describing the case where a standard FA is carried out separately in each study.
The observed variables in study $s$ are decomposed into $T_s$ factors. 
In particular, let $\mbl_{is},\;i=1,\dots, n_s$ be the values of the \textit{study-specific} factors in individual $i$ of study $s$ and
$\bsL_s,\, s=1,\dots, S$ be the $P \times T_s$ corresponding factor loading matrix.
FA assumes that $\mbx_{is}$ is decomposed as
\begin{eqnarray}
\mbx_{is} &=&  \bsL_s \mbl_{is} + \mbe_{is} \;\;\;\;i=1,\dots,n_s \, ,
\label{fa}
\end{eqnarray}
where $\mbe_{is}$ is a  normal
error term with covariance matrix $\bsP_s = \diag (\psi_{1s}, \dots, \psi_{ps})$ \citep[e.g.][]{joreskog1967, joreskog1967some, joreskog1972factor}.
FA aims at explaining the dependence structure among  observations
by decomposing the $P \times P$ covariance matrix $\bsS_s$ as
$\bsS_s = \bsL_s \bsL_s^\top + \bsP_s \, . $

\begin{figure}[h]
\centering
\includegraphics[width=14.5cm]{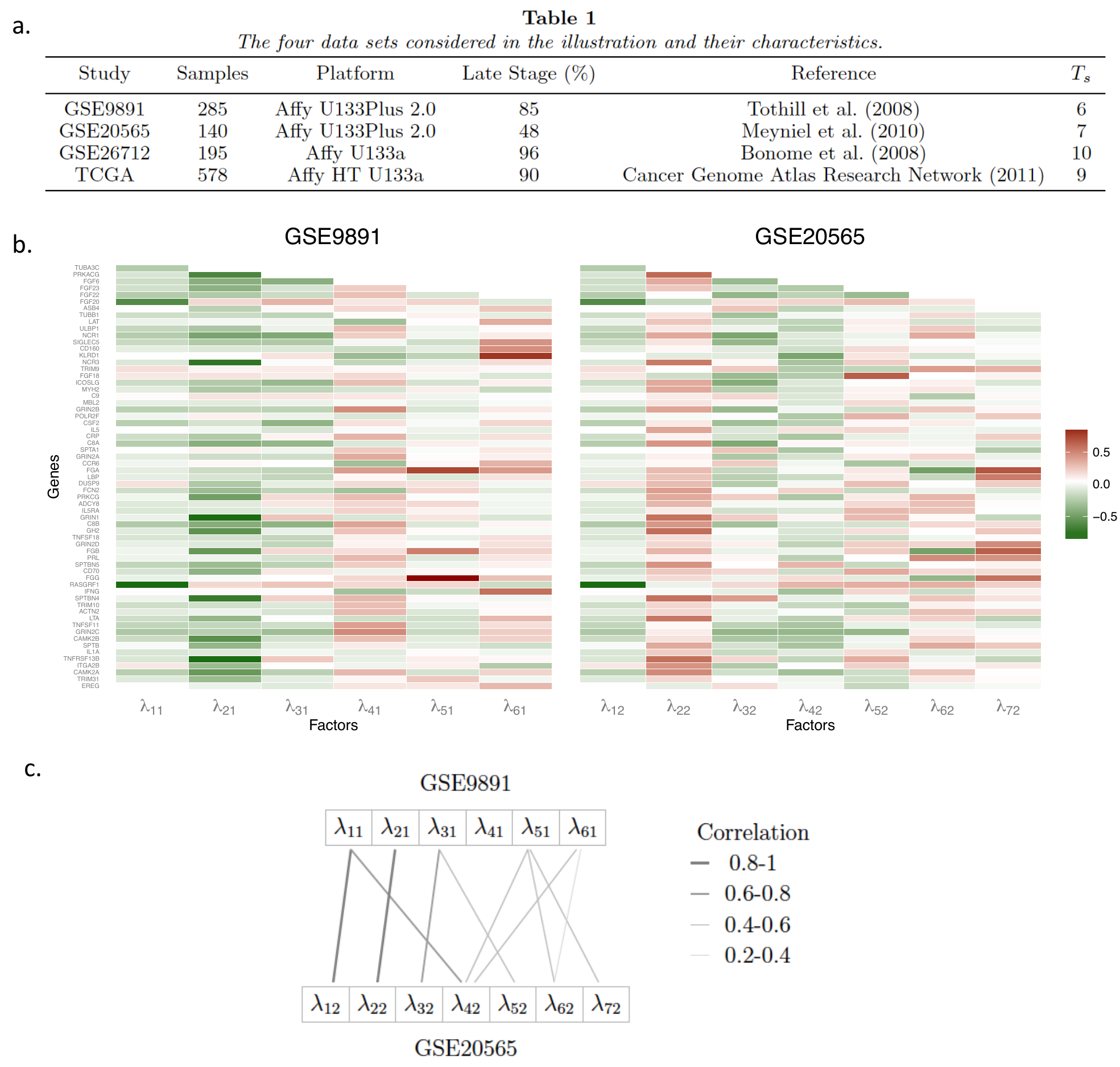}
\caption{\textbf{a}: Table of the four data sets considered in the analysis. It provides an overview of the studies with corresponding samples, information about the specific microarray platform used, the proportion of patients diagnosed with Stage III or Stage IV OC, references, and finally the total latent dimension $T_s$ obtained by \texttt{nFactors} package in \texttt{R}. \textbf{b}: Heatmap of the estimated factor loadings, $\bsL_1 = \left\{\bsl_{11},\bsl_{21},\bsl_{31}, \bsl_{41}, \bsl_{51},\bsl_{61}\right\}$ for the GSE9891 study and $\bsL_2 = \left\{\bsl_{12},\bsl_{22},\bsl_{32},\bsl_{42},\bsl_{52},\bsl_{62}, \bsl_{72}\right\}$,  obtained by performing a separate factor analysis as in equation (\ref{fa}) in studies GSE9891 and GSE20565 from~Figure 1.a. Each column $\bsl_{is}$ is thus the  $i^{th}$ loading vector of the $s^{th}$ study.  We estimate parameters by maximum likelihood estimation (MLE), implemented via the Expectation-Maximization (EM) algorithm using the identifying constraint that the loading matrix is lower triangular, as will be detailed in  \S \ref{sub:iden}-\ref{sub:est}.
\textbf{c}: Graphical representation of the absolute value of correlations between pairs of study-specific factor loadings.  Correlations smaller than .2 are not shown. Darker lines denote larger correlations in absolute value. In these correlations 
the same variables are considered in each study, and their order is preserved. 
\label{fig: esempioFA} }
\label{fig: mfa_cor}
\end{figure}

Figure~\ref{fig: esempioFA}.a  shows an overview of the studies   analyzed in this work for which more information can be found in Supplementary Materials \S A.   %

Figure~\ref{fig: esempioFA}.b suggests that some
 loading vectors may reveal common patterns across studies. 
 This point is further explored in Figure~\ref{fig: esempioFA}.c where  three of the  loading vectors of the
GSE9891 study are strongly correlated
with four loading vectors in the GSE20565 study. 
Highly correlated pairs of loading vectors are more likely to represent common factors. On the other hand, some loading vectors of GSE9891 (e.g. $\bsl_{41}$) exhibit low correlation with all loading vectors of GSE20565. These loadings are likely to result from feature unique to this study. 
Furthermore, in order to assess the multivariate correlation between the two loading matrices, we compute the RV coefficient \citep{robert1976}. It is equal to 0.76 showing a very similar pattern between the two matrices, and equal to 0.86 taking in consideration the two matrices with the first four factor loadings only.
  
The MSFA model proposed here is designed to analyze multiple studies jointly, replacing the heuristic interpretation above with a principled statistical approach.
It explicitly models common biological features shared among the studies, as well as
 unique variation present in each study. Specifically, the observed variables in study $s$ are decomposed into $K$ factors shared with the other studies, and $J_s$ additional factors reflecting its unique sources of variation, for a total of $T_s = K+J_s$ factors. 
 Let $\mbf_{is}$ be the \textit{common} factor vector in subject $i$ of study $s$, and 
$\bsPH$ be the $P \times K$ common factors loading matrix. Moreover, let $\mbl_{is}$ be the \textit{study-specific} factor and
$\bsL_{s}$ be the $P \times J_s$ specific factors loading matrix.
MSFA assumes that the $P$-dimensional centered response vector $\mbx_{is}$ can be written as
\begin{equation}
\mbx_{is} = \bsPH \mbf_{is} + \bsL_s \mbl_{is} + \mbe_{is} \,,
\qquad i=1, \dots, n_s
\qquad s=1,\dots, S.
\label{eqn:MFA}
\end{equation}
where the $P \times 1$ random error vector $\mathbf{e}_{is}$ has a multivariate normal distribution with mean vector $\mathbf{0}$ and covariance matrix 
 $\mathbf{\bsP}_{s}$, with $\mathbf{\bsP}_{s} = \diag (\psi_{\textit{s}_1}^2, \dots, \psi_{\textit{s}_p}^2)$. 
We also assume that the marginal distribution of $\mathbf{l}_{is}$ is multivariate normal with mean 
vector $\mathbf{0}$ and covariance matrix  $\mathbf{I}_{J_s}$, and  the marginal distribution of  $\mathbf{f}_{is}$ is multivariate normal with mean
vector  $\mathbf{0}$ and covariance matrix $\mathbf{I}_{k}$, where $\mathbf{I}$ denotes the identity matrix.

As a result of the model
 assumptions, the marginal distribution of $\mathbf{x}_{is}$ is multivariate normal with mean vector $\mathbf{0}$ and covariance matrix $\bsS_s = \bsPH \bsPH^\top+\bsL_s\bsL_s^\top+\bsP_s$, with the three
terms reflecting the variance of the common factors, the variance of the study-specific factors, and the variance of the error, 
respectively.

This model can be applied to many settings when 
the aim is to isolate commonalities from differences across groups, population or studies. 
In the gene expression application of interest here
the focus is on estimating the biological signal shared among the studies, while removing study-specific features less likely to be reproducible across populations, and potentially arising from technological issues. Elsewhere the goal may be to capture some study-specific features of interest while removing common factors shared among the studies.  Finally,  other applications may focus on 
 both common and specific factors.  Different examples focused in the study-specific pattern could be find in nutritional epidemiology \citep{carrera2007study, ryman2015} where population-specific diets have a lower or higher impact in specific disease, such as obesity and cancer.

\subsection{Identifiability}
\label{sub:iden}

To specify an identifiable model, the MSFA model must be further constrained to avoid orthogonal rotation indeterminacy, similarly to the classic FA. 
Let  $\bsOm_s = [\bsPH, \bsL_s]$ be the $P \times (K+J_s)$
 loading matrix for the $s^{th}$ study. If we define $\bsOm_s^\ast=\bsOm_s \mathbf{Q}_s$,
where   $\mathbf{Q}_s$ is a 
square orthogonal matrix with $(K+J_s)$ rows, it readily follows that $\bsOm_s^\ast (\bsOm_s^\ast)^\top= \bsOm_s \mathbf{Q}_s \mathbf{Q}_s^\top
\bsOm_s^\top=\bsOm_s \bsOm_s^\top$, so that 
$
\bsS_s =\bsOm_s^\ast (\bsOm_s^\ast)^\top + \mathbf{\bsP}_s=\bsOm_s \bsOm_s^\top  + \mathbf{\bsP}_s \, ,
$
and $\bsS_s$ is not uniquely identified. 

FA (\ref{fa}) identifies the parameters by imposing constraints on the factor loadings matrix.
One possibility often used in practice is to take $\bsL_s$ in 
(\ref{fa})
as a lower triangular matrix \citep{Geweke_1996, Lopes_2004,  carvalho2008}.  
Here we adapt  this approach to the MSFA, and   specify
 $\bsOm_s = [\bsPH, \bsL_s]$ 
to be lower triangular. Importantly,
 the  matrices $\bsPH$ and $\bsL_s$ are not interchangeable.
As in  FA, assuming a lower triangular form for $\bsOm_s$ resolves the orthogonal rotation indeterminacy \citep[][pp. 565-566]{Geweke_1996}.  There remain the issue that
we can simultaneously change the sign to all the elements of the loading matrices and to all the 
latent factors without  changing the model. This could be fixed by constraining the sign of a subset of loadings, but for MLE of the model parameters
this issue is largely inconsequential.  

An important issue in maximum likelihood estimation of the MSFA model parameters concerns the complexity
of the specifications that can be handled by the method, since there are some constraints that ought to be
considered. In particular, for the $s^{th}$ study, the number of elements in the sample covariance matrix, $\mbC_{x_s x_s}$, 
 must be no greater than the number of parameters in $\bsS_s$,
 implying that 
$ P (K + J_s) + P - \frac{(K+J_s) (K+J_s - 1)}{2}  \leq \frac{1}{2}P (P+1) \, , s=1,\ldots,S \, .$ 
Another important constraint  is that 
effective application of 
 the method  essentially requires more observations than variables. 
We  will  initially assume  that $P < \min\left\{ n_1, \dots, n_s\right\}$.
Some possible directions to consider to overcome the limitations
will be highlighted in the Discussion.


\subsection{Parameter estimation}
\label{sub:est}
The parameters to be estimated in the MSFA are $\bsth = \left(  \bsPH, \bsL_s,  \bsP_s \right)$.  For notational simplicity in both (\ref{fa}) and (\ref{eqn:MFA}) we assume that  the observed variables in each study have
been centered. 
The log-likelihood  function corresponding to the MSFA assumptions 
is
$$
\ell (\bsth)= \log \prod_{s=1}^{S}  \prod_{i=1}^{n_s} p(\mbx_{is}\vert \bsth)  = \sum_{s=1}^{S} \left\{ -\frac{n_s}{2} \, \log| \bsS_s| - \frac{n_s}{2} {\rm tr}(\bsS_s^{-1}\,  \mbC_{x_sx_s}) \right\} \,.
$$

In the following, the MLE will be obtained by means of the  Expectation Conditional Maximization (ECM) algorithm  \citep{MENG_1993}, a class of generalized EM algorithms \citep{Dempster_77, rubin1982}. The details of the ECM algorithm for the MSFA model are reported in 
the Supplementary Materials.


\subsubsection{Dimension Selection}

Selecting the dimension of the model can be challenging.  
The following two-step procedure was found to be effective in the settings of interest.
First  the total latent dimension $T_{s} = K+J_s$ for each of the $S$ studies
is determined by
 using  standard
  techniques for FA, such  as Horn's parallel analysis  \citep{horn1965}, Cattell's scree test \citep{cattell1966} 
 or the use of  indexes, such as the RMSEA \citep{steiger1980}.  Next, model selection techniques 
 are applied to the 
  overall MSFA model to 
 select the number $K$ of latent factors sharing a common loading matrix $\bsPH$, as described in \S  \ref{sub:sele}. The dimensions
 $J_s$ are then obtained residually as $T_s - K$, with the restrictions that $T_s - K \geq 0$ for all $s=1,\ldots,S$.

\section{Simulation studies}
 
 We perform simulation experiments to evaluate the effectiveness  of the  ECM algorithm 
in  estimating  the MSFA model parameters, as well as the suggested strategy for selecting the dimensionality of the latent factors.
 The simulation studies  are designed to closely mimic the data 
of Figure~\ref{fig: esempioFA}.a,  therefore   $S=4$ studies are considered, with 
dimension of the latent factors $T_s = \left\{ 6, 7, 10, 9\right\}$,  and 
$\mbx_{is}$ generated  from $P$-dimensional normal distributions, with sample size equal to $n_s =\left\{ 285,140,195,578 \right\}$. The samples from 
the various 
studies are assumed to be drawn from the different population, each with zero mean and covariance matrix $\bsS_s = \bsPH \bsPH^\top+\bsL_s\bsL_s^\top+\bsP_s, \,\mbox{ with } \, S=1,\dots,4$.

 Three simulation scenarios are investigated.
In  Scenario 1 there are no common factors, i.e. $K = 0$, in
 Scenario 2  $K = 1$, and in  Scenario 3 $K = 3$. 
To achieve more realistic results, 
  in each scenario the data are generated by parameter values  akin to those estimated with the data introduced in Section 2.
 
 \subsection{Parameter estimation via the ECM algorithm}
 
We first analyze the performances of the ECM algorithm for a given selection of $K$ and $J_s$, $s=1,\ldots,S$. 

Irrespective of the optimization method adopted, 
 the  choice of the starting point   is  crucial for  achieving good performances. 
 The
  following strategy is proposed, for given
  factor dimensions $K$ and $J_s$, $s=1,\ldots,S$.
 \begin{enumerate}
\item  A single data set is created by stacking the data of the four studies by row,  obtaining a single data set with  $n = n_1 + n_2 + n_3 + n_4=1198$ observations and $P=100$ variables.
\item A  Principal Components Analysis (PCA) is performed on the  data obtained at step (1). The first $K$ components are
taken as the starting point of the common factor loadings. 
\item We perform FA to the separated data.
The loadings and uniquenesses obtained from the FA are used as the initial values for $\bsL_s$ and $\bsP_s$.
\end{enumerate}

 \begin{figure}[h]
\centering
\resizebox{1\textwidth }{!}{ %
\includegraphics[width=30cm]{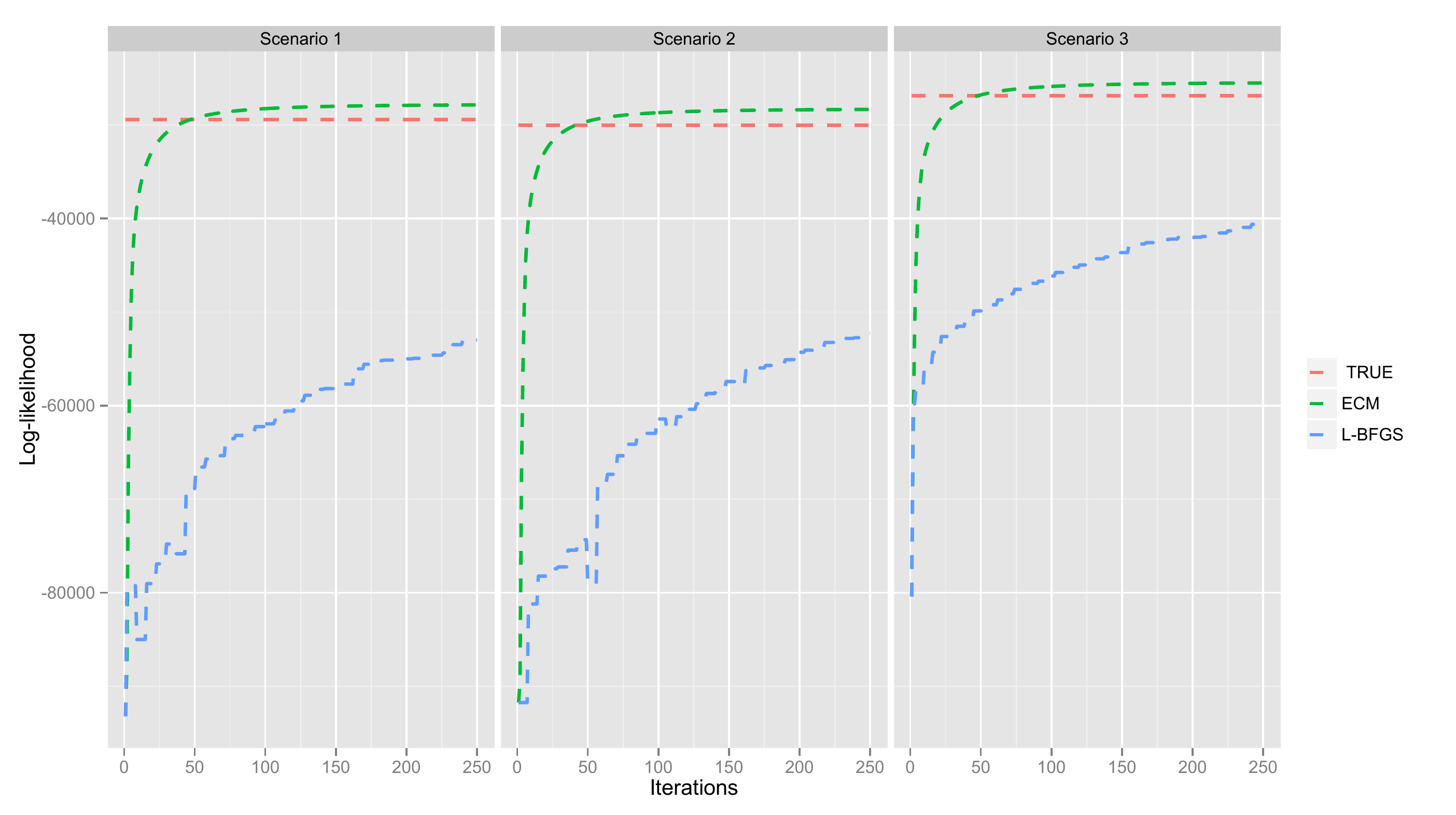}
}
\caption{Log-likelihood by iteration. This graph compares the ascent of the log-likelihood function obtained with the ECM algorithm (green line)  to those obtained
by maximizing the log-likelihood $\ell(\bsth)$ by a standard quasi-Newton optimizer, given by 
the box-constrained Limited-memory Broyden-Fletcher-Goldfarb-Shanno 
 (L-BFGS) method \citep[e.g.][]{byrd1995} (blue line),
whereas the red line represents the log-likelihood calculated at the true parameter value. Each panel refers to a single representative data set, for each of the three scenarios.  The two algorithms start from the same point. 
This method is used
 for benchmarking the ECM algorithm in its standard form, without any particular effort to tailor the optimization to the model at hand. }
\label{fig: 3lik}
\end{figure}

Figure \ref{fig: 3lik} shows that the convergence of the ECM algorithm is obtained with fewer number of iterations compared to  box-constrained Limited-memory Broyden-Fletcher-Goldfarb-Shanno 
 (L-BFGS) method \citep[e.g.][]{byrd1995}. 
 Moreover,  
  severe convergence problems 
 arise when the L-BFGS method is 
used  for larger number   $P$ of variables.

\begin{figure}[!h]
\centering
\resizebox{1\textwidth }{!}{ %
\includegraphics[width=14.5cm]{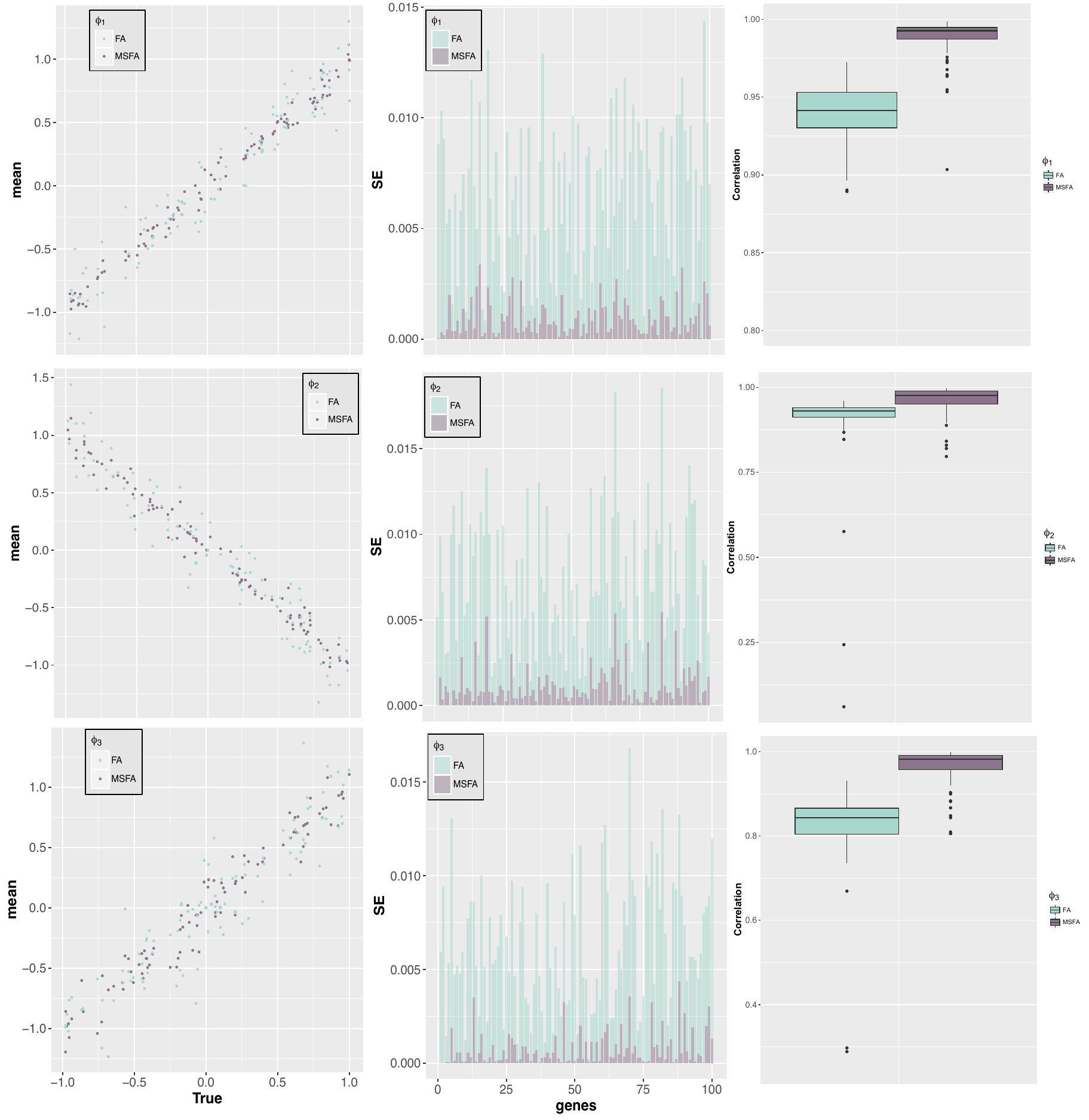}
}
\caption{Distribution of the common factor loadings estimated by MSFA (in purple) and FA (in green) from 100 simulations under Scenario 3, i.e. $K=3$. The graph compares the distribution of the estimated three factor loadings identified by the MSFA (in purple) and FA (in green) after stacking the datasets in one.  The left panel shows the mean of the estimated factor loadings after the 100 simulation and the distance from the true common factor loadings. Comparison with FA are also reported. In the center we report the standard errors of these distances. Standard errors of each variable or gene obtained by MSFA (in purple) are smaller than those obtained by FA (in green) for all the three common factor estimates. Finally, in the right panel we report the boxplot showing the correlation between the estimated factor loadings and the true common loadings for each simulation.}
\label{fig: mean_k3}
\end{figure}

Figure \ref{fig: mean_k3} shows that  MSFA is able to recover the true common factor loadings. Moreover, MSFA performs better than FA in terms of estimating the true shared factor loadings. FA is computed after stacking the studies in a unique dataset.   Different analysis for checking if the MSFA recovers the true factors, and results for the other scenarios are reported in Supplementary Materials.

\subsection{Selection of the latent factor dimensions}
\label{sub:sele}
The selection of 
 the dimension of the latent space  is studied, again by means of simulation
experiments performed under the same three  scenarios considered above.  For each data set, we first choose $T_s$ by means of standard FA techniques, 
and then choose $K$.  We thus focus
in particular on the problem of selecting $K$, tackled by
 applying existing model selection techniques,  such as the  
  AIC and BIC criteria, for which there is an extensive  literature 
 \citep{burnham2002, BIC}. 
 The study of model selection based on information criteria  is still in progress in  FA settings \citep{chen2008extended, hirose2014estimation},  so it seems useful to   evaluate the behavior of both AIC and BIC  for choosing $K$.
 Along the two information-based criteria,  the  likelihood ratio test (LRT) for choosing between nested models 
 with different values of $K$ is considered as well.

Table~\ref{tab: aic2} shows the results obtained by our model fitting approach for 100 different data sets generated independently from
the MSFA under the three different scenarios.

\begin{table}[ht]
\captionof{table}{Comparison of model assessment methods under  Scenario 1, 2 and 3. Results are obtained by our
model fitting approach on simulations for 100 different data sets generated independently from
the MSFA with $K=0$, $K=1$ and $K=3$. In Scenario 1 the overall trend is that 
all the three methods favor choosing the model with $K = 0$, but 
both AIC and LRT do so more consistently than BIC. In Scenario 2 and 3 AIC outperforms both  BIC and LRT, though the latter is not far off.  }
\label{tab: aic2}
\centering
\begin{tabular}{c c c c c c c c c}
\hline\hline
 & Method & \textcolor{red}{$K = 0$} & $K = 1$ & $K = 2$ & $K = 3$ & $K = 4$ & $K = 5$ \\ [0.5ex] 
\hline
 & AIC & 100 & 0 & 0 & 0 & 0 & 0  \\
 Scenario 1 & BIC & 91 & 1 & 2 & 6 & 0 & 0 \\
 & LRT & 100 & 0 & 0 & 0 & 0 & 0\\
\hline\hline
& Method & $K = 0$ &  \textcolor{red}{$K = 1$} & $K = 2$ & $K = 3$ & $K = 4$ & $K = 5$ \\ [0.5ex] 
\hline
 &AIC & 0 & 100 & 0 & 0 & 0 & 0  \\
Scenario 2 & BIC & 0 & 0 & 0 & 2 & 6 & 92 \\
&LRT & 3 & 97 & 0 & 0 & 0 & 0\\
 [1ex]
\hline\hline 
& Method & $K = 0$ & $K = 1$ & $K = 2$ &  \textcolor{red}{$K = 3$} & $K = 4$ & $K = 5$ \\ [0.5ex] 
\hline
& AIC & 0 & 0 & 0 & 100 & 0 & 0  \\
 Scenario 3 & BIC & 0 & 0 & 0 & 1 & 23 & 76 \\
& LRT & 0 & 0 & 0 & 91 & 9 & 0\\
 [1ex]
\hline
\end{tabular}
\end{table}

AIC  seems to be the best criterion, always leading to the selection of the
true model. The contrast with the performance of the BIC is striking.  Generally, BIC penalizes model complexity more strongly than  AIC, so it is not surprising that BIC tends to prefer models with more common factor loadings and thus less parameters. In our application, subtracting a unit to $K$ adds a large number of parameters to the model, as more factors are allowed to differ across studies.

The results of these three simulation studies  point strongly towards the usage of AIC to select
the value of $K$. This will therefore be the strategy employed in our applied example.

\section{Gene expression example: Immune System pathway}

To illustrate MSFA in an important biological example, we analyzed the four studies described in Figure~\ref{fig: esempioFA}.a where the sample size $n_s (s=1, \dots, 4)$ are listed, focusing on transcription of genes involved in immune system activity ($P=63$). Specifically, we considered genes included in the sub-pathways ``Adaptive Immune System'' (AI), 
``Innate Immune System'' (II) and
``Cytokine Signaling in Immune System'' (CSI) from {\tt reactome.org}. These sub-pathways belong to the larger Immune System pathway and do not have overlapping genes. In addition, we restricted attention to genes which are common across all studies.

Initially, we conducted preliminary analyses to asses the total latent factor dimensions, the number of common factors across studies and the number of specific factors for each study. 
Using the AIC, the number of common factors is set to one.

We then compare the cross-validation prediction errors computed by the MSFA to those computed by FA. Notice the latter method is applied in two different ways, namely the first is merging 4 studies into one single data set and the second is to separately compute FA in each study. We fit the MSFA and the standard FA on a random $80$\% of the data, and evaluate the prediction error on the remaining $20\%$. Predictions are obtained as
\[
\mbox{MSFA:} \;\;\;\;\;\hat{\mbx}_{is} =  \hat{\bsPH} \hat{\mbf}_{i} + \hat{\bsL}^{MSFA}_{s}\hat{\mbl}_{is} \;\;\;\;\;\;\;\;\;\;\;\;\;\;\;\;\;\;\;\;\;\;\;\;
\mbox{FA:} \;\;\;\;\; \hat{\mbx}_{is} =  \hat{\bsL}^{FA}_{s}\hat{\mbl}_{is}.
\]
where $\hat{\bsL}^{MSFA}_{s}$ are the specific factor loadings estimated with MSFA and $\hat{\bsL}^{FA}_{s}$ are the factor loadings estimated with FA.
We evaluated the mean squared error of prediction
$
\mbox{MSE} = \frac{1}{n}\sum_{s=1}^S \sum_{i=1}^n \left( x_{is} - \hat{x}_{is}\right)^2,
$
using the $20$\% of the samples in each study set aside for each cross-validation iteration. The MSE is 4\% smaller for MSFA than for FA after merging the data and is 0.048\% smaller for MSFA than for FA applied separately to each study.

This analysis illustrates how MSFA borrows strength across studies in the estimation of the factor loadings, in such a way that the predictive ability in independent observation is not only preserved but even improved. Moreover, the AIC obtained with FA after stacking all the study in a dataset is higher than the AIC computed by MSFA.

\begin{figure}[!h]
\centering
\resizebox{1\textwidth }{!}{ %
\includegraphics[width=14.5cm]{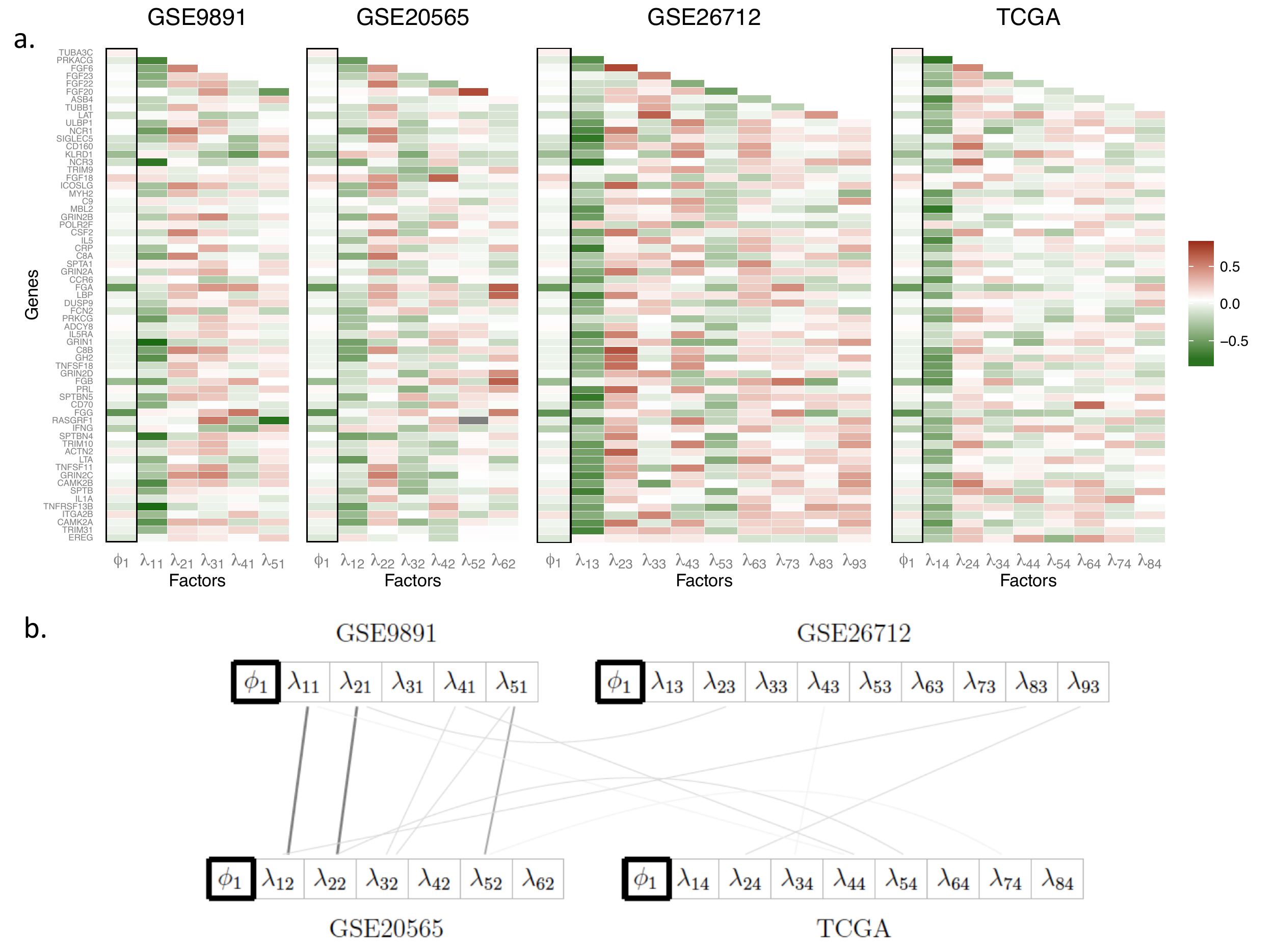}
}
\caption{\textbf{a.}Heatmap of the estimated factor loadings obtained with MSFA, both common (black rectangle) and specific ones performed in the data sets in Figure~\ref{fig: esempioFA}.a . \textbf{b.} Graphical representation of the cross-study pairwise correlation of the specific factor loadings obtained with the MSFA. Darker grey lines correspond to higher correlations. Correlations smaller than .25 are not shown. Absolute correlations range from 0.66 to 0.81.}
\label{fig: 4study}
\end{figure}

Next, we focus on the analysis of the factor themselves. 
The heatmap in Figure~\ref{fig: 4study}.a depicts the estimates of the factor loadings, both the common factor (highlighted in the black rectangle) that can be identified reproducibly across the studies, and the specific ones.

To help interpreting the biological meaning of the common factor, we apply Gene Set Enrichment Analysis (GSEA) for determining whether one of the three gene sets is significantly enriched among loadings that are high in absolute value \citep{mootha2003, subramanian2005}.  We consider all the three sub-pathways in the Immune System pathway.
We used the package \texttt{RTopper} in \texttt{R} in \texttt{Bioconductor}, following the method illustrated in \citet{tyekucheva2011}. The resulting analysis shows that the common factor is significantly enriched by the II system sub-pathway, suggesting that genuine biological signal may have been identified.

Further,  Figure \ref{fig: mfa_cor}.b shows that three of the specific factors of the GSE9891 study are strongly correlated with three corresponding factors in the GSE20565 study.

To further probe this possibility, at least within the Immune System pathway, we analyze studies GSE9891 and 
GSE20565 separately from the other two using MSFA (Figure reported in Supplementary materials). 
The AIC chooses a model with $K=4$. Studies GSE9891 and GSE20565 use the same microarray platform, Affy~U133~Plus2.0, unlike the other two. This prompts the conjecture that the three stronger correlations observed may be related to technological rather than biological variation. Naturally it is also possible that there may be specific technical features of this platform that enable it to identify additional factors, although this is less likely in view of the fact that our analysis is restricted to a common set of genes. 

We performed the GSEA on the estimated factor loadings for the two-study analysis. 
The results show that the first common factor is still related to the II system pathway, as was the case for the single common factor shared between the four studies in the earlier analysis. Also, the common factor of the four studies analysis is highly correlated with the first common factors of the two-study analysis, $r=0.60$. 
The three remaining common factors are not related to any of the remaining pathways, further corroborating the hypothesis that they may represent the results of spurious variation unique to the specific platform used.

We also checked the impact of the choice of a gene order, because of the dependence induced by the block lower triangular structure
 assumed for $\bsOm_s$, to address identifiability.
In particular, we repeated the same analysis after permuting the variables.
 Despite minor discrepancies,  the final conclusion is not changed. Namely, the
 single common factor is significantly enriched only with the innate immune system
 sub-pathway.

Overall, this analysis illustrates important features of this method, including its ability to capture biological signal common to multiple studies and technological platforms, and at the same time to isolate the source of variation coming, for example, from the different platform by which gene expression is measured. 

\section{Discussion}

In this article we introduced and studied a novel class of factor analysis methodologies for the joint analysis of multiple studies. We have provided a valuable tool to accelerate the pace at which we can combine unsupervised analysis across multiple studies, and understand the cross-study
reproducibility of signal in multivariate data.

The main concept is to separately identify and estimate 1) common factors shared across multiple studies, and 2) study-specific factors. This is intended to help address one of the most critical steps in cross-study analysis, namely
to identify factors that are reproducible
across studies and to remove idiosyncratic variation that lacks cross-study reproducibility. The method is simple and it is based on a generalized version of FA able to handle multiple studies simultaneously and to capture the two types of information.

Several methods have been proposed to analyze diverse data sets and to capture the correlation between different studies. The CPCA was introduced by \cite{flury1984} to investigate the hypothesis that the covariance matrices for different populations are simultaneously diagonalizable. This method estimates a common principal axes across the different population and the deviation of the data from the model of common principal axes.
The Co-inertia analysis (CIA) emerged in ecology to explore the common structure of two distinct sets of variables (such as species' abundances of flora and fauna) measured at the same sites \citep{doledec1994, dray2003}. It proceeds by separately performing dimension reduction on each set of variables, to derive factor scores for the sites. In a second, independent, stage the correlation between these factors is investigated. The Multiple Co-inertia analysis (MCIA) \citep{dray2003} is a generalization of CIA to consider more than two data sets.  MCIA finds a hyperspace, where variables showing similar trends are projected close  to  each other \citep{meng2014}. 
 
A related  method is the Multiple Factor analysis (MFA) \citep{abdi2013}, an extension of principal component analysis (PCA)  which  consists  of three steps. The first is a PCA  for each study. In the second step each data set is normalized by dividing by its first singular value. In the third step, a single data set is created by stacking the normalized data from different studies by row, and a final PCA is done. 


Two differences can be emphasized between these approaches and Multi-study Factor Analysis. First they are focused on analyzing only  the common structure after having excluded the noise. Instead our method gives an estimation of both common and study-specific components. Second they operate stage-wise, decomposing each matrix separately, while our study analyzed the data jointly. This is critical in a meta-analytic context because the presence of a recognizable factor in one study can assist with the identification of the same factor in other studies even when it is more difficult to recognize it.

We develop a fast ECM algorithm for parameter estimates and  
 provide a procedure for choosing the common and specific factor. 
The MSFA needs to be constrained to be identifiable and so the constraints used here is the popular 
block lower triangular matrix. Although this condition is largely used in classical FA settings, it 
 induces an order dependence among the variables \citep{fruhwirth2010}. As noted in 
  \cite{carvalho2008},  the  choice of the first 
$K+J_s$ variables is an important modeling decision, to be made with some care.  In our application, it is somewhat reassuring that 
the checking made on the impact of the chosen variable order on the final conclusion leads to the same conclusions, although general conclusions cannot be drawn. Others constraints or rotation methods, such as the varimax criterion \citep{kaiser1958}, could be considered, though their
extension to the MSFA setting would require some further investigation.

In  settings characterized by high-dimensional data,  the $n>P$ condition  requires   variations to  the  proposed computations.
To this end, our ongoing research is focusing on the extension of   the  Bayesian infinite factor model proposed 
by  \citet{bhattacharya2011}  to the MSFA setting.

The MSFA model can be applied to many settings when 
the aim is to isolate commonalities and differences across different groups, population or studies. There might be other applications where the goal is to 
 capture study-specific features of interest and, instead, remove common factors shared among studies. 
 Other applications may focus on capturing both common and specific factors, without removing any of them. 
MSFA may have broad applicability in a wide variety of genomic platforms (e.g. microarrays,
RNA-seq, SNPs, proteomics, metabolomics, epigenomics), as well as datasets in other fields of biomedical
research, such as those generated by exposome studies or Electronic Medical Record (EMR). However, the concept is straightforward, universal and of general interest across all applications of multivariate analysis.

An R package, namely \textit{MSFA}, is implemented and available on GitHub at \url{https://github.com/rdevito/MSFA}.



\clearpage

\bibliographystyle{biom}
\bibliography{refsRDV1}

\end{document}